\documentclass[letterpaper, 10 pt, conference]{ieeeconf}  % Comment this line out
\usepackage{amsmath,amsfonts,amssymb}                   \usepackage{graphicx} 
\usepackage{multirow}
\usepackage{color}
\usepackage{setspace}
\usepackage{tikz}
\usepackage{lipsum,adjustbox}
\usetikzlibrary{calc,positioning}
\usetikzlibrary{decorations.pathreplacing,calligraphy}
\usepackage{multicol}
\usepackage{threeparttable}
\usepackage{comment}
\usepackage[acronym]{glossaries}
\usepackage{amsfonts}
\usepackage{balance}

\usepackage{marginnote}

\newacronym{saf}{saf}{Store-and-forward}
\newacronym{ST}{ST}{Service Station}
\newacronym{LWR}{LWR}{Lighthill-Whitham-Richards}
\newacronym{CTM-s}{CTM-s}{Cell Transmission Model with service station}
\newacronym{CTM}{CTM}{Cell Transmission Model}
\newacronym{CTT}{CTT}{Constant Travel Time}
\newacronym{METANET-s}{METANET-s}{METANET with service station}

\newacronym{VSL}{VSL}{Variable Speed Limit}

\usepackage{mathtools}
\usepackage{amsmath}

\newcommand{\R}{\mathbb{R}}
\newcommand{\N}{\mathbb{N}}
\newcommand{\bs}[1]{\boldsymbol{#1}}

\newcommand{\tup}[1]{\textup{#1}}

\IEEEoverridecommandlockouts                              % This command is only
\title{\LARGE \bf
A new control-oriented METANET model\\ to encompass service stations on highways
}

\author{Ayda Kamalifar, Carlo Cenedese, Michele Cucuzzella, Antonella Ferrara% <-this % stops a space
\thanks{Ayda Kamalifar and Antonella Ferrara acknowledge  support from PNRR-M4C2-I1.4-NC-HPC-Spoke6.
The work of Carlo Cenedese is supported by NCCR Automation and funded by the Swiss National Science Foundation (grant number 180545).}% <-this % stops a space
\thanks{Ayda Kamalifar, Michele Cucuzzella, and Antonella Ferrara are with the Department of Electrical, Computer Science and Biomedical Engineering, Univeristy of Pavia, 27100 Pavia, Italy
        {\tt\small (ayda.kamalifar01@universitadipavia.it, michele.cucuzzella@unipv.it,
        antonella.ferrara@unipv.it})}%
        \thanks{Carlo Cenedese is with Department of Information Technology and Electrical Engineering, ETH Z\"urich, Z\"urich, Switzerland 
        {\tt\small (ccenedese@ethz.ch)}}%
}

\begin{document}

\maketitle
\thispagestyle{empty}
\pagestyle{empty}

%%%%%%%%%%%%%%%%%%%%%%%%%%%%%%%%%%%%%%%%%%%%%%%%%%%%%%%%%%%%%%%%%%%%%%%%%%%%%%%%
\begin{abstract}
In this paper, we propose the \gls{METANET-s} model, a second-order macroscopic traffic model that, compared to the classical METANET, incorporates the dynamics of service stations on highways.  
Specifically, we employ the (so-called) store-and-forward links to model the stop of vehicles and the possible queue forming in the process of merging back into the highway mainstream. 
We explore the capability of the \gls{METANET-s} to capture well both traffic back propagation and capacity drops, which are typically caused by the presence of vehicles joining again the mainstream traffic from the service station. Therefore, capturing these effects is crucial to improving the model's predictive capabilities. Finally, we perform a comparative analysis with the \gls{CTM-s}, showcasing that the \gls{METANET-s} describes the traffic evolution much better than its first-order counterpart.
\end{abstract}

%%%%%%%%%%%%%%%%%%%%%%%%%%%%%%%%%%%%%%%%%%%%%%%%%%%%%%%%%%%%%%%%%%%%%%%%%%%%%%%%

\section{INTRODUCTION}
Nowadays, transportation experts face the complex challenge of balancing the urgency for sustainable solutions and the rising traffic demand due to the growing urbanization. There is pressing need for a transition in modern mobility that mitigates emissions, time inefficiency, and unproductive fuel consumption~\cite{angelidou2022emerging}. Traffic models and simulations play a fundamental role in addressing these challenges \cite{kotsialos2001importance}. They empower experts to attain a profound understanding of the real issues within transportation systems, identify opportunities for enhancement, predict and measure the impacts of infrastructure developments~\cite{siri2021freeway}. Essentially, these tools enable informed decision-making and control systems contributing to the endeavor for a more sustainable and efficient transportation network~\cite{ganin2017resilience}.

Macroscopic first-order traffic models have a long and rich history, starting from the seminal \gls{LWR} model \cite{lighthill1955kinematic}, and the \gls{CTM}, which corresponds to the discretized version of the \gls{LWR}~\cite{daganzo1995cell}. 
% In the \gls{CTM}, traffic flow is expressed as a function of density, and speed is assumed not to vary dynamically across different sections of the roadway.
The \gls{CTM} has been successfully applied in many applications, including simulation of large-scale motorway networks \cite{lebacque1996godunov}, dynamic traffic assignment \cite{lebacque1996strada}, and traffic estimation \cite{alecsandru2011assessment}. A recent noteworthy extension of the \gls{CTM} is the  \gls{CTM-s}, firstly proposed in~\cite{cenedese2022novel,cenedese2022optimal}. 
Specifically, the \gls{CTM-s} endows the classical \gls{CTM} with the dynamics necessary to model the presence of a \gls{ST}. The rapid grow of electric vehicles creates a shift in the use of \glspl{ST} by drivers. These facilities provide not only the possibility of charging and/or refueling but also improved ancillary services that have the effect of increasing the time spent by drivers during a stop~\cite{cenedese2021highway}. Moreover, policymakers aim to have a service station every 60 km in the whole EU by 2030~\cite{fitfor55,cenedese:2022:incentive_EV_bottleneck}. Therefore, the \gls{CTM-s} has been proven to be a useful tool for planning~\cite{cenedese2022optimal} and studying the effect of \glspl{ST} on the evolution of traffic~\cite{cenedese2022novel,cenedese:2023:capacity_drops}. 
Nonetheless, first-order traffic models, including \gls{CTM-s}, exhibit one main limitation.
Due to the lack of knowledge on the traffic velocity, these models are not able to capture complex traffic phenomena such as capacity drops or have to rely on convoluted tricks to somehow capture them~\cite{cenedese:2023:capacity_drops}.  

To overcome this limitation, higher-order models can be adopted. In~\cite{messner1990metanet}, the METANET model is introduced to describe also the speed evolution,  enabling to capture instabilities such as stop-and-go waves, and capacity drops.
In~\cite{cremer1981parameter,fan2013data}, the METANET has been shown to describe the traffic flow much better than first-order models. In \cite{wang2020freeway,wang2022real}, further refinements have been made to the METANET model to improve traffic state estimation and prediction. Finally,  control strategies like model predictive control, ramp-metering, and \gls{VSL} have been explored in \cite{ferrara2021hierarchical,carlson2010optimal,carlson2010optimall}.

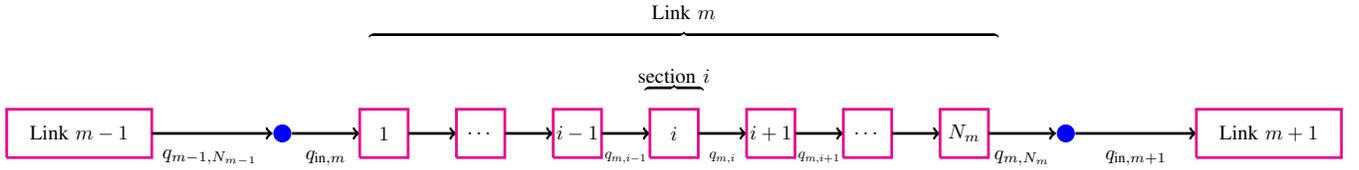
\begin{figure*}
\centering
\begin{adjustbox}{width=1\textwidth}
\begin{tikzpicture}
\draw[ultra thick] [magenta] (-9,-0.5) -- (-9,0.5) -- (-6,0.5) -- (-6,-0.5) -- (-9,-0.5);
\draw[ultra thick] [->] (-6,0) -- (-3.5,0);
\filldraw[blue] (-3.3,0) circle (5pt) node[]{};
\draw[ultra thick] [->] (-3.1,0) -- (-1.7,0);
\draw[ultra thick][magenta] (-1.7,-0.5) -- (-1.7,0.5) -- (-0.7,0.5) -- (-0.7,-0.5) -- (-1.7,-0.5);
\draw[ultra thick] [->] (-0.7,0) -- (0.3,0);
\draw[ultra thick][magenta] (0.3,-0.5) -- (0.3,0.5) -- (1.3,0.5) -- (1.3,-0.5) -- (0.3,-0.5);
\draw[ultra thick] [->] (1.3,0) -- (2.3,0);
\draw[ultra thick][magenta] (2.3,-0.5) -- (2.3,0.5) -- (3.3,0.5) -- (3.3,-0.5) -- (2.3,-0.5);
\draw[ultra thick] [->] (3.3,0) -- (4.3,0);
\draw[ultra thick][magenta] (4.3,-0.5) -- (4.3,0.5) -- (5.3,0.5) -- (5.3,-0.5) -- (4.3,-0.5);
\draw[ultra thick] [->] (5.3,0) -- (6.3,0);
\draw[ultra thick][magenta] (6.3,-0.5) -- (6.3,0.5) -- (7.3,0.5) -- (7.3,-0.5) -- (6.3,-0.5);
\draw[ultra thick] [->] (7.3,0) -- (8.3,0);
\draw[ultra thick][magenta] (8.3,-0.5) -- (8.3,0.5) -- (9.3,0.5) -- (9.3,-0.5) -- (8.3,-0.5);
\draw[ultra thick] [->] (9.3,0) -- (10.3,0);
\draw[ultra thick][magenta] (10.3,-0.5) -- (10.3,0.5) -- (11.3,0.5) -- (11.3,-0.5) -- (10.3,-0.5);
\draw[ultra thick] [->] (11.3,0) -- (12.7,0);
\filldraw[blue] (12.9,0) circle (5pt) node[]{};
\draw[ultra thick] [->] (13.1,0) -- (15.6,0);
\draw[ultra thick] [magenta] (15.6,-0.5) -- (15.6,0.5) -- (18.6,0.5) -- (18.6,-0.5) -- (15.6,-0.5);

\node[scale=1.2] at (-7.5,0) {Link $m-1$};
\node[scale=1.2] at (-1.2,0) {$1$};
\node[scale=1.2] at (0.8,0) {$\cdots$};
\node[scale=1.2] at (2.8,0) {$i-1$};
\node[scale=0.9] at (3.8,-0.5) {$q_{m,i-1}$};
\node[scale=1.2] at (4.8,0) {$i$};
\node[scale=0.9] at (5.8,-0.5) {$q_{m,i}$};
\node[scale=1.2] at (6.8,0) {$i+1$};
\node[scale=0.9] at (7.8,-0.5) {$q_{m,i+1}$};
\node[scale=1.2] at (8.8,0) {$\cdots$};
\node[scale=1.2] at (10.8,0) {$N_m$};
\node[scale=1.2] at (17.1,0) {Link $m+1$};
\draw[ultra thick][decorate,decoration={calligraphic brace}] (-1.5,2) -- (11.5,2);
\node[scale=1.2] at (5,2.5) {Link $m$};
\draw[ultra thick][decorate,decoration={calligraphic brace}] (4.2,0.9) -- (5.4,0.9);
\node[scale=1.2] at (4.8,1.2) {section $i$};
%\draw[ultra thick][decorate,decoration={calligraphic brace}] (-9.2,2) -- (-5.8,2);
%\node[scale=1.2] at (-7.5,2.5) {link $m-1$};
%\draw[ultra thick][decorate,decoration={calligraphic brace}] (15.4,2) -- (18.8,2);
%\node[scale=1.2] at (17.3,2.5) {link $m+1$};

\node[scale=1.2] at (-4.8,-0.5) {$q_{m-1,N_{m-1}}$};
\node[scale=1.2] at (-2.4,-0.5) {$q_{\text{in},m}$};
\node[scale=1.2] at (12,-0.5) {$q_{m,N_{m}}$};
\node[scale=1.2] at (14.35,-0.5) {$q_{\text{in},m+1}$};
\end{tikzpicture}
\end{adjustbox}%
%\end{center}
\caption{Link discretization of a freeway stretch in METANET model}
\label{linkdescritization}
\end{figure*}

The main contributions of this paper are the following: \textit{i)}  we develop the \gls{METANET-s}, a novel highway traffic network model that is capable of capturing the dynamics of \glspl{ST} on highways; 
\textit{ii)} we incorporate the physical constraints of  \glspl{ST} into the model, ensuring that the number of vehicles at a \gls{ST} does not exceed the \gls{ST} capacity; \textit{iii)} we use fundamental concepts of the classical METANET ensuring that the model can be easily implementable by practitioners to predict traffic evolution and design decision strategies; \textit{iv)} we show via numerical simulations that the \gls{METANET-s} can describe complex phenomena such as capacity drops much better than the \gls{CTM-s}.

%%%%%%%%%%%%%%%%%%%%%%%%%%%%%%%%%%%%%%%%%%%%%%%%%%%%%%%%%%%%%%%%%%%%%%%%%%%%%%%%

\section{Preliminaries on the classical METANET model}
\label{sec2}
In this section, we introduce the key components of the classical METANET model as a cornerstone over which we build in the next section the proposed \gls{METANET-s} model. In the following we adopt a similar notation as in~\cite{pasquale2018new}.
 
The METANET model discretizes time into intervals of length $T\in\R$ and indexed by $k\in\N$. The highway stretch is divided into $N\in\N$ \textit{links}. Each link $m\in\bs N\coloneqq\{1,\dots,N\}$ is divided into $N_{m}$ \textit{sections} composed of $\lambda_{m}$ lanes of length $L_{m}$, which are used to better describe the variation of the link features. The set of all the sections of link $m$ is denoted by $\bs N_{m}\coloneqq\{1,\dots,N_m\}$; see Fig. \ref{linkdescritization}. Every two adjacent links in $\bs N$ are connected by \textit{nodes} (blue dots in Fig. \ref{linkdescritization}). We denote the total number of nodes by $P$ and the corresponding set by $\bs P\coloneqq\{1,\dots,P\}$. For each section $i\in \bs N_{m}$ of link $m\in \bs N$, the following variables are defined:
 \begin{itemize}
     \item \textit{density} $\rho_{m,i}(k)$ [veh/km lane], the number of vehicles in the section $i$ during the time interval $k$ for each lane $\lambda_{m}$ and divided by $L_{m}$; 
     \item \textit{mean speed} $v_{m,i}(k)$ [km/h] of the vehicles moving along the section $i$;
     \item \textit{flow} $q_{m,i}(k)$ [veh/h lane], the number of vehicles exiting section $i$ during $k$ and divided by $T$ for each lane $\lambda_{m}$. 
 \end{itemize}
The METANET model can operate in two different modes: the {\textit{non destination-oriented}}, and the {\textit{destination-oriented}}.
In the former, the traffic assignment---specifically the behavior of drivers choosing their routes---is not taken into account, while in the latter it is~\cite{ferrara2018freeway}. In the upcoming section, we employ the destination-oriented operation mode to distinguish  between the behavior of vehicles that stop at the \gls{ST} and those that do not.

To properly define the destination-oriented mode, the following additional variables are introduced:
\begin{itemize}
    \item \textit{partial traffic density} $\rho_{m,i,j}(k)$  [veh/km lane], the density of vehicles in $i\in\bs N_m$ with destination $j\in J_{m}$, where $J_{m}\subseteq\bs N$ is the set of destination links that can be accessed from link $m\in\bs N$;
    \item \textit{composition rate} $\gamma_{m,i,j}(k)\in[0,1]$, is the fraction of traffic with destination $j\in{J_{m}}$. 
\end{itemize}
Using the METANET model, the whole transportation network is described as a collection of nodes, describing junctions or bifurcations, connected by links. For the readers' convenience, we briefly introduce in the following sections the dynamics associated to the METANET links and nodes (see e.g.~\cite{kotsialos1998modelling} for further details).

\subsection{The links}  
The links in the METANET can be of four different  types serving specific purposes and endowed with unique features. We discuss them hereafter.

\subsubsection{Freeway links} are employed for homogeneous freeway stretches where the dynamics read as
\begin{subequations}
\begin{align} \label{equation1}
 \rho_{m,i}(k+1)&=\rho_{m,i}(k)+\frac{T}{L_{m}\lambda_{m}}\bigg[q_{m,i-1}(k)
 -q_{m,i}(k)\bigg]\\     
 \label{equation2}
   q_{m,i}(k)&=\rho_{m,i}(k) v_{m,i}(k) \lambda_{m} \\
   \label{equation3}
     v_{m,i}(k+1)&=v_{m,i}(k)+\frac{T}{\tau} \nonumber \bigg[v(\rho_{m,i}(k))-v_{m,i}(k)\bigg]\\ \nonumber
    &\:\: +\frac{T}{L_{m}}v_{m,i}(k)\bigg[v_{m,i-1}(k)-v_{m,i}(k)\bigg]\\ 
    &\:\: -\frac{VT[\rho_{m,i+1}(k)-\rho_{m,i}(k)]}{\tau L_{m}[\rho_{m,i}(k)+K]}\\
\label{equation4}
    v(\rho_{m,i}(k))&\coloneqq \bar v_{m}\,\exp\bigg[-\frac{1}{a_{m}}\left(\frac{\rho_{m,i}(k)}{\rho_{\textup{cr},m}}\right)^{a_{m}}\bigg],
\end{align}
\end{subequations}
where \eqref{equation1} is derived from the conservation principle and \eqref{equation2} from the definition of the flow, while \eqref{equation3} empirically describes the average speed evolution.
% speed relation that determines speed evolution for each segment $i$ of link $m$. 
The function defined in \eqref{equation4} describes the desired speed given the current density of section $i$. A brief description of the parameters in \eqref{equation1}-\eqref{equation4} is reported in Table~\ref{tab1}.
Extra terms can be incorporated into \eqref{equation3} to model lane reductions and merging phenomena near on-ramps, see \cite{kotsialos2002traffic}. Moreover, the partial densities for each destination $j\in J_m$ read as
\begin{subequations} \label{equation5}
\begin{align}
 &\rho_{m,i,j}(k+1)=\rho_{m,i,j}(k)\nonumber  \\
 &\hspace{.5cm}+\frac{T}{L_{m}\lambda_{m}} \bigg[\gamma_{m,i-1,j}(k) q_{m,i-1}(k)-\gamma_{m,i,j}(k)q_{m,i}(k)\bigg] \label{equation2a}  \\ 
&\gamma_{m,i,j}(k)=\frac{\rho_{m,i,j}(k)}{\rho_{m,i}(k)}. \label{equation2b}
\end{align}
\end{subequations}
In the case that $i=1$, the composition rate and flow entering the first section of link $m$ is equal to the composition rate and the flow exiting the previous link, i.e., $\gamma_{m-1,N_{m-1},j}(k)$ and $q_{m-1,N_{m-1}}(k)$, respectively.
\begin{table}[t]
\caption{Static data defined in METANET model.} 
\label{tab1}
\begin{center}
 \begin{tabular}{|c||c||c|}
 \hline
 \textbf{Parameters} & \textbf{Descriptions} & \textbf{Units}\\
 \hline
  $\bar v_{m}$   & \footnotesize{Free-flow speed} & \textup{[km/h]}\\
  \hline
 $\rho_{\textup{cr},m}$ &   \footnotesize{Critical density per lane} & \textup{[veh/km lane]} \\
 \hline
 $a_m$ & \footnotesize{Exponent parameter} & \textup{[-]}\\
 \hline
 $\tau$    & \footnotesize{Relaxation time} & \textup{[h]}\\ 
 \hline
 $V$ &  \footnotesize{Anticipation constant} & [${\text{km}}^2$/\textup{h}]\\
  \hline
  $K$   &  \footnotesize{Numerical stability parameter} & \textup{[veh/km]}\\
   \hline
   $\delta$   &  \footnotesize{Merging parameter} & \textup{[h/km]}\\
   \hline
   $\phi$   &  \footnotesize{Lane drop parameter} & \textup{[h/km]}\\
   \hline
 \end{tabular}   
\end{center} 
% \vspace{-12mm}
\end{table}

\subsubsection{\gls{saf} links} differently from the freeway links, they include neither density and speed dynamics nor link discretization. They are queue models represented by their maximum capacity $q_{\textup{max},s}(k)$, queue length $w_{s}(k)$. For each \gls{saf} link $s\in\bs N^{\tup{saf}}\subseteq \bs N$, the incoming flow of vehicles creates a queue. The evolution of the queue length is attained via the conservation principle, i.e.,
\begin{equation}\nonumber
   w_{s}(k+1)=w_{s}(k)+T\left[q_{\textup{in},s}(k)-q_{s}(k)\right]
\end{equation} 
Then, the queuing vehicles are directed to the succeeding downstream link after a certain time delay. The dynamics of the outflow for $s\in\bs  N^{\tup{saf}}$ are as follow
\begin{equation}
    q_{s}(k)=\min\bigg[q_{\textup{in},s}(k)+\frac{w_{s}(k)}{T},q_{\textup{max},s}(k)\bigg],
\end{equation}
where $q_{\textup{in},s}(k)$ [veh/h] is the entering flow and $q_{\textup{max},s}(k)$ is the maximum outflow of the \gls{saf} link which  depends on the density of the downstream link $d\in\bs N$ and is defined as
\begin{equation}\nonumber
    q_{\textup{max},s}(k)= 
\begin{cases}
  Q_{s} & \text{if $\rho_{d,1}(k)<\rho_{\textup{cr,d,1}}$} \\
  Q_{s}P(k) & \text{otherwise},
\end{cases}
\end{equation}
where $Q_{s}$ is the maximum flow of $s\in\bs  N^{\tup{saf}}$ and $P(k)$ is the portion of it that can enter link $d$. This fraction is defined as $P(k)\coloneqq\frac{\rho_{\textup{max},d,1}-\rho_{d,1}}{\rho_{\textup{max},d,1}-\rho_{\textup{cr},d,1}}$, where $\rho_{\textup{max},d,1}, \rho_{\textup{cr},d,1}$ and $\rho_{d,1}$ are, respectively, the jam density, critical density and actual density of the first section of the downstream link $d$, hence discriminating whether the whole demand can enter or not.

In the destination-oriented mode, the notion of the partial queues $w_{s,j}(k)$ for each destination {$j\in J_s$} reachable from link $s$ has to be introduced, thus
\begin{align}\nonumber
    w_{s,j}(k+1)=&~w_{s,j}(k)\\
    &+T\left[\gamma_{\textup{in},s,j}(k)q_{\textup{in},s}(k)  -\gamma_{s,j}(k)q_{s}(k)\right],\nonumber
\end{align}
where $\gamma_{\textup{in},s,j}(k)$ is the \textit{composition rate} describing the portion of flow entering the \gls{saf} link $s$ and having {$j\in J_{s}$} as destination and $\gamma_{s,j}(k)=\frac{w_{s,j}(k)}{w_{s}(k)}$ indicates the distribution of the flow $q_{s}(k)$ exiting the \gls{saf} link $s$ with destination {$j\in J_{s}$}.

 \subsubsection{Origin links}serve as entry points for the traffic demand $d_o(k)\in\R$. Each origin link $o\in \bs N$ is modelled as a \gls{saf} link \cite{kotsialos2002traffic}.

 \subsubsection{Destination links} are the exit points for the traffic flow. Traffic conditions in these links are strongly influenced by the conditions in the next link and are assumed to be not congested if there is no available measurement.

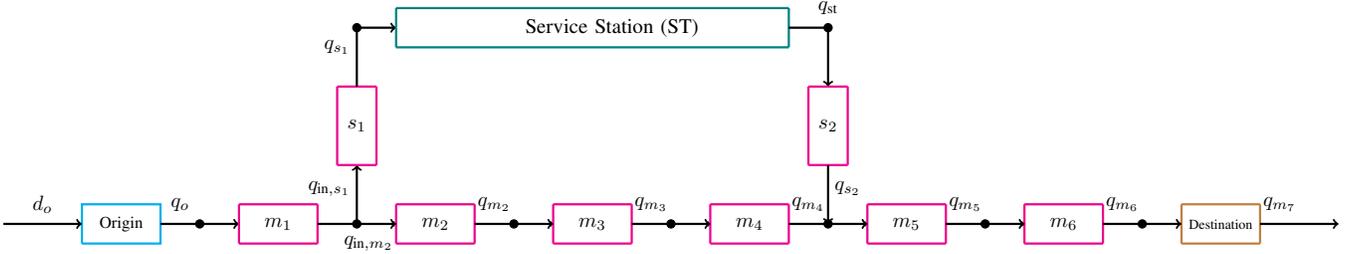
\begin{figure*}
\centering
\begin{adjustbox}{width=\textwidth}
\begin{tikzpicture}

%origin link
\draw[ultra thick][cyan] (0,0) -- (2,0) -- (2,1) -- (0,1)-- (0,0);
\draw[ultra thick] [->] (-2,0.5) -- (0,0.5);
%mainstream
\draw[ultra thick] [->] (2,0.5) -- (4,0.5);
\draw[ultra thick][magenta] (4,0.5) -- (4,1);
\draw[ultra thick][magenta] (4,1) -- (6,1);
\draw[ultra thick][magenta] (6,1) -- (6,0);
\draw[ultra thick][magenta] (6,0) -- (4,0);
\draw[ultra thick][magenta] (4,0) -- (4,0.5);
\draw[ultra thick][->] (6,0.5) -- (8,0.5);
\draw[ultra thick][magenta] (8,0.5) -- (8,1);
\draw[ultra thick][magenta] (8,1) -- (10,1);
\draw[ultra thick][magenta] (10,1) -- (10,0);
\draw[ultra thick][magenta] (10,0) -- (8,0);
\draw[ultra thick][magenta] (8,0) -- (8,0.5);
\draw [ultra thick][->](10,0.5) -- (12,0.5);
\draw[ultra thick][magenta] (12,0.5) -- (12,1);
\draw[ultra thick][magenta] (12,1) -- (14,1);
\draw[ultra thick][magenta] (14,1) -- (14,0);
\draw[ultra thick][magenta] (14,0) -- (12,0);
\draw[ultra thick][magenta] (12,0) -- (12,0.5);
\draw[ultra thick][->] (14,0.5) -- (16,0.5);
\draw[ultra thick][magenta] (16,0.5) -- (16,1);
\draw[ultra thick][magenta] (16,1) -- (18,1);
\draw[ultra thick][magenta] (18,1) -- (18,0);
\draw[ultra thick][magenta] (18,0) -- (16,0);
\draw[ultra thick][magenta] (16,0) -- (16,0.5);
\draw[ultra thick][->] (18,0.5) -- (20,0.5);
\draw[ultra thick][magenta] (20,0.5) -- (20,1);
\draw[ultra thick][magenta] (20,1) -- (22,1);
\draw[ultra thick][magenta] (22,1) -- (22,0);
\draw[ultra thick][magenta] (22,0) -- (20,0);
\draw[ultra thick][magenta] (20,0) -- (20,0.5);
\draw[ultra thick][->] (22,0.5) -- (24,0.5);
\draw[ultra thick][magenta] (24,0.5) -- (24,1);
\draw[ultra thick][magenta] (24,1) -- (26,1);
\draw[ultra thick][magenta] (26,1) -- (26,0);
\draw[ultra thick][magenta] (26,0) -- (24,0);
\draw[ultra thick][magenta] (24,0) -- (24,0.5);
\draw[ultra thick][brown] (28,1) -- (30,1);
\draw[ultra thick][brown] (30,1) -- (30,0);
\draw[ultra thick][brown] (30,0) -- (28,0);
\draw[ultra thick][brown] (28,0) -- (28,0.5);
\draw[ultra thick][brown] (28,0.5) -- (28,1);
\draw[ultra thick][->] (30,0.5) -- (32,0.5);
%st
\draw[ultra thick] [->](7,0.5) -- (7,2);
\draw[ultra thick][magenta] (7,2) -- (7.5,2);
\draw[ultra thick][magenta] (7,2) -- (6.5,2);
\draw[ultra thick][magenta] (7.5,2) -- (7.5,4);
\draw[ultra thick][magenta] (6.5,2) -- (6.5,4);
\draw[ultra thick][magenta] (6.5,4) -- (7.5,4);
\draw [ultra thick][-](7,4) -- (7,5.5);
\draw[ultra thick] [->](7,5.5) -- (8,5.5);
\draw [ultra thick][teal](8,5.5) -- (8,6);
\draw [ultra thick][teal](8,5.5) -- (8,5);
\draw [ultra thick][teal](8,6) -- (18,6);
\draw [ultra thick][teal](18,6) -- (18,5);
\draw [ultra thick][teal](18,5) -- (8,5);
\draw [ultra thick][-](18,5.5) -- (19,5.5);
\draw [ultra thick][->](19,5.5) -- (19,4);
\draw [ultra thick][magenta](19,4) -- (19.5,4);
\draw [ultra thick][magenta](19,4) -- (18.5,4);
\draw [ultra thick][magenta](19.5,4) -- (19.5,2);
\draw [ultra thick][magenta](18.5,4) -- (18.5,2);
\draw [ultra thick][magenta](18.5,2) -- (19.5,2);

%add text
\node[scale=1.2] at (1,0.5) {\textup{Origin}};
\node[scale=1.5] at (-1,1) {$d_{o}$};
\node[scale=1.5] at (5,0.5) {$m_{1}$};
\node[scale=1.5] at (7,3) {$s_{1}$};
\node[scale=1.5] at (19,3) {$s_{2}$};
\node[scale=1.5] at (13.5,5.5) {\textup{Service Station (ST)}};
\node[scale=1.5] at (9,0.5) {$m_{2}$};
\node[scale=1.5] at (13,0.5) {$m_{3}$};
\node[scale=1.5] at (17,0.5) {$m_{4}$};
\node[scale=1.5] at (21,0.5) {$m_{5}$};
\node[scale=1.5] at (25,0.5) {$m_{6}$};
\node[scale=1] at (29,0.5) {\textup{Destination}};

\node[scale=1.5] at (2.5,1) {$q_{o}$};
\node[scale=1.5] at (18.5,1) {$q_{m_{4}}$};
\node[scale=1.5] at (22.5,1) {$q_{m_{5}}$};
\node[scale=1.5] at (26.5,1) {$q_{m_{6}}$};
\node[scale=1.5] at (30.5,1) {$q_{m_{7}}$};
\node[scale=1.5] at (7.3,0) {$q_{\tup{in},m_{2}}$};
\node[scale=1.5] at (6.3,1.4) {$q_{\tup{in},s_{1}}$};
\node[scale=1.5] at (10.5,1) {$q_{m_{2}}$};
\node[scale=1.5] at (14.5,1) {$q_{m_{3}}$};
\node[scale=1.5] at (19.5,1.4) {$q_{s_{2}}$};
\node[scale=1.5] at (6.5,5) {$q_{s_{1}}$};
\node[scale=1.5] at (19,6) {$q_{\tup{st}}$};

%arrows
%\draw [->] (14,5.5) -- (15,5.5);
%\draw [->] (15,5.5) -- (15,4);
%\draw [->] (15,2) -- (15,0.5);
\draw [->] (15,0.5) -- (16,0.5);
\draw [->] (18,0.5) -- (20,0.5);
\draw [->] (22,0.5) -- (24,0.5);
\draw [->] (6,0.5) -- (8,0.5);
\draw [->] (10,0.5) -- (12,0.5);
\draw[ultra thick] [->] (19,2) -- (19,0.5);
\draw[ultra thick] [->] (26,0.5) -- (28,0.5);
\filldraw[black] (3,0.5) circle (3pt) node[]{};
\filldraw[black] (7,0.5) circle (3pt) node[]{};
\filldraw[black] (11,0.5) circle (3pt) node[]{};
\filldraw[black] (15,0.5) circle (3pt) node[]{};
\filldraw[black] (19,0.5) circle (3pt) node[]{};
\filldraw[black] (23,0.5) circle (3pt) node[]{};
\filldraw[black] (27,0.5) circle (3pt) node[]{};
\filldraw[black] (7,5.5) circle (3pt) node[]{};
\filldraw[black] (19,5.5) circle (3pt) node[]{};
\end{tikzpicture}
\end{adjustbox}%
%\end{center}
\caption{Depiction of the \gls{METANET-s} model. The freeway links (pink rectangles) represent the mainstream and on-ramp/off-ramp of service station which are connected by the nodes (black dots), an origin link (blue rectangle), a destination link (brown rectangle), and the \gls{ST} is modeled via a \gls{saf} link (green rectangle).}
\label{networksample}
\end{figure*}

\subsection{The nodes} At locations corresponding to junctions, bifurcations, merging on-ramps, and diverging off-ramps, nodes play an essential role in representing the interconnections between various links within the network. For each node $p\in{\bs P}$, the set of entering and exiting links are denoted by $I_{p}\subseteq \bs N$ and  $O_{p}\subseteq \bs N$, respectively.
Then, $Q_{p}(k)$ is the total traffic flow entering node $p$ during $k$. Similarly, the flow leaving $p$ through a link $d\in O_{p}$ is denoted by $q_{\textup{in},d}(k)$.
Additionally, the \textit{turning rate} $\beta_{p}^d(k)$, indicates the portion of traffic flow $Q_{p}(k)$ which leaves node $p$ at period $k$ via link $d\in O_{p}$. 
The relation among these quantities is described as follows:
\begin{subequations}
\begin{align}
Q_{p}(k)=\sum_{m \in I_{p}}{q_{m,N_{m}}}(k) 
    \label{equation11}\\
q_{\textup{in},d}(k)=\beta_{p}^d(k)Q_{p}(k),  
    \label{equation12}    
\end{align}
\end{subequations}
where $q_{m,N_{m}}(k)$ is the flow exiting the last section of link $m\in I_p$ and $q_{\textup{in},d}(k)$ that entering the first section of the exiting link $d$.
In the destination-oriented mode, multiple splitting rates $\beta_{p,j}^d(k)$ replace the single turning rate $\beta_{p}^d(k)$. Let $Q_{p,j}(k)$ be the total traffic flow entering node $p$ and destined to {$j\in J_{p}$}, which is a destination link reachable from node $p$. Then, the splitting rate $\beta_{p,j}^d(k)$ is the portion of the flow $Q_{p,j}(k)$ leaving $p$ through link $d\in O_{p}$. For every node $p\in\bs P$ with destination {$j\in J_{p}$} the following hold: 
\begin{subequations}
\begin{align}
Q_{p,j}(k)&=\sum_{m \in I_{p}}{q_{m,N_{m}}(k) \gamma_{m,N_{m},j}(k)} 
\label{equation13}\\
q_{\textup{in},d}(k)&=\sum_{{j\in J_{p}}}{\beta_{p,j}^d(k) Q_{p,j}(k)}
\label{equation14}\\
\gamma_{\textup{in},d,j}(k)&=\frac{\beta_{p,j}^d(k) Q_{p,j}(k)}{q_{\textup{in},d}(k)}.
\label{equation15}
\end{align}
\end{subequations}
Note that \eqref{equation13}--\eqref{equation15} provide $\gamma_{m,1,j}(k)$ and $q_{m,1}(k)$ used in~\eqref{equation5} when $i=1$.
%%%%%%%%%%%%%%%%%%%%%%%%%%%%%%%%%%%%%%%%%%%%%%%%%%%%%%%%%%%%%%%%%%%%%%%%%%%%%%%%%%%%%%%%%%%%%%%%%%%%%%%%%%%%%%%%%%%%%%%%%%%%%%%%%%%%%%%%%%%%%%%%%%%%%%%%%%%%%%%%%%%%%%%%%%%%%%%%%%%%%%%%%%%%%%%%%%%%%%%%%%%%%%%%%%%%%%%%%%%%%%%%%%%%%%%%%%%%%%%%%%%%%%%%%%%%%%%%%%%%%%%%%%%%%%%%%%%%%%%%%%%%%%%%%%%%%%%%%%%%%%%%%%%%%%%%%%%%%%%%%%%%%%%%%%%%%%%%%%%%%%%%%%%%%%%%%%%%%%%%%%%%%%%%%%%%%%%%%%%%%%%%%%%%%%%%%%%%%%%%%%%%%%%%%%%%%%%%%%%%%%%%%%%%%%%%%%%%%%%%%%%%%%%%%%%%%%%%%%%%%%%%%%%%%%%%%%%%%%%%%%%%%%%%%%%%%%%%%%%%%%%%%%%%%%%%%%%%
%\newpage
%\addtolength{\textheight}{-3cm}   % This command serves to balance the column lengths

%%%%%%%%%%%%%%%%%%%%%%%%%%%%%%%%%%%%%%%%%%%%%%%%%%%%%%%%%%%%%%%%%%%%%%%%%%%%%%%%
\section{The METANET with service station (\gls{METANET-s})}
 
In this section, the classical METANET model is endowed with the dynamics of a \gls{ST}. Figure~\ref{networksample} is a visual aid for the structure of the  \gls{METANET-s} and the used notation.
The freeway stretch is partitioned into $N$ freeway links, interconnected by $P$ nodes. Without loss of generality, we suppose that each link is subdivided into only one section, i.e., $N_{m}=1$, $\forall{m\in\bs N}$. An origin link $o\in\bs N$ is used to inject into the network the traffic demand $d_{o}(k)\in\R$. Vehicles can exit the mainstream through the link $s_{1}\in\bs N$ to reach the \gls{ST}, which is represented by a \gls{saf} link. A queue of vehicles stay for a certain amount of time $\delta$ at the \gls{ST} to use ancillary services such as resting and refueling. After spending $\delta$ amount of time at the \gls{ST}, drivers attempt to merge back into the mainstream. Then, if the flow exiting the \gls{ST}, i.e., $q_{\textup{st}}(k)$, exceeds the capacity $q_{\textup{max,st}}(k)$, a queue of vehicles forms at the exit point of the \gls{ST}.
Thus, a coupling should be considered between the number of vehicles entering the \gls{ST} and the number of  vehicles attempting to exit in the next time intervals. %Since the \gls{ST} is defined as a \gls{saf} link, the former dynamics is covered by the dynamics of the \gls{saf} link. But the letter dynamics should be defined particularly. Consequently, this situation distinctly impacts the flow, density, and speed of vehicles attempting to merge back to the freeway stretch. The \gls{METANET-s} model explicitly addresses and models these dynamics. 
Furthermore, a constraint should be considered for the entering flow to the \gls{ST} in order to avoid traffic congestion in the entry point. 

\subsection{\gls{METANET-s} dynamics} 
The origin link $o\in\bs N$ receives as input the traffic demand $d_{o}(k)$ and thus the exiting flow is defined as
\begin{equation} \label{equation16}
   q_{o}(k) =r_{o}(k)\min\left[d_{o}(k)+\frac{w_{o}(k)}{T},q_{\textup{max},o}(k)\right], 
\end{equation}
where $r_{o}(k)$ is the metering rate for the link $o$ (if it is not applied, then $r_{o}(k)=1$ for all $k\in\N$). The maximum flow of the origin link depends on the density of the first section of the first downstream link, i.e.,  $\rho_{m_{1}}(k)$, hence

\begin{equation}
   q_{\textup{max},o}(k) =
   \begin{cases}
  Q_{o} & \text{if $\rho_{m_{1}}(k)<\rho_{\textup{cr},m_{1}}$} \\
  Q_{o}P(k) & \text{otherwise}.
  \label{equation17}
\end{cases}
\end{equation}
The permit function $P(k)\coloneqq\frac{\rho_{\textup{max},m_{1}}-\rho_{m_{1}}(k)}{\rho_{\textup{max},m_{1}}-\rho_{\textup{cr},m_{1}}}$ influences the outflow of the origin link $q_{o}(k)$ based on the congestion level of the exiting link $m_{1}$. If link $m_{1}$ is dense or congested, the capacity of the origin link $q_{\textup{max},o}(k)$ is reduced. As the capacity decreases, a queue of vehicles forms at the origin link $o$, resulting in
\begin{equation}
   w_{o}(k+1)=w_{o}(k)+T\bigg[d_{o}(k)-q_{o}(k)\bigg].
\end{equation}
Density and speed dynamics for link $m_{1}$ are introduced similarly to \eqref{equation1}-\eqref{equation4}. In order to identify the portion of the flow $q_{m_{1}}(k)$ entering the mainstream, i.e., $q_{\textup{in},m_{2}}(k)$, and the \gls{ST}, i.e., $q_{\textup{in},s_{1}}(k)$, it is necessary to consider partial densities as in \eqref{equation2a}, i.e., 
    \begin{align}
    &\rho_{m_{1},s_{1}}(k+1)=~\rho_{m_{1},s_{1}}(k)\nonumber\\
    &\hspace{.4cm}+\frac{T} {L_{m_{1}}\lambda_{m_{1}}} \bigg[\gamma_{o,s_{1}}(k)q_{o}(k) 
  -\gamma_{m_{1},s_{1}}(k)q_{m_{1}}(k)\bigg] \\
 &\rho_{m_{1},m_{2}}(k+1)=~\rho_{m_{1},m_{2}}(k)\nonumber\\
 &\hspace{.4cm}+\frac{T} {L_{m_{1}}\lambda_{m_{1}}} \bigg[\gamma_{o,m_{2}}(k)q_{o}(k) -\gamma_{m_{1},m_{2}}(k)q_{m_{1}}(k)\bigg],
    \end{align}
where $\gamma_{o,s_{1}}(k)$, $\gamma_{m_{1},s_{1}}(k)$, $\gamma_{o,m_{2}}(k)$, and $\gamma_{m_{1},m_{2}}(k)$ represent the composition rates at time period $k$ based on  \eqref{equation2b} and satisfy the following constraints
\begin{subequations}
    \begin{align}
        \gamma_{o,s_{1}}(k)+\gamma_{o,m_{2}}(k)&=1 \nonumber\\
        \gamma_{m_{1},s_{1}}(k)+\gamma_{m_{1},m_{2}}(k)&=1. \nonumber
    \end{align}
\end{subequations}
The vehicles entering the \gls{ST} through the  link $s_{1}$ during $k$, stop at the \gls{ST} for  $\delta$ amount of time before trying to merge back into the mainstream. Then, the number of vehicles at the \gls{ST} evolves as
\begin{equation}
\ell_{\textup{st}}(k+1)=\ell_{\textup{st}}(k)+T\bigg[q_{s_{1}}(k)-q_{\textup{st}}(k)\bigg],
\end{equation}
where, $q_{s_{1}}(k)$ and $q_{\textup{st}}(k)$ denote the entering and exiting flow of the \gls{ST} in the time interval $k$, respectively.
After $\delta$ time intervals, vehicles attempt to exit the \gls{ST} to merge back into the mainstream. However, if the flow  $q_{\textup{st}}(k)$ exceeds the maximum capacity $q_{\textup{max,st}}(k)$, a queue forms at the exit point of the \gls{ST}, including vehicles waiting for merging back into the mainstream in the next time intervals. Let $w_{\textup{st}}(k)$ denote the queue length, its dynamics can be expressed as
\begin{equation}
    w_{\textup{st}}(k+1)=w_{\textup{st}}(k)+T\bigg[q_{s_{1}}(k-\delta)-q_{\textup{st}}(k)\bigg].
\end{equation}
Then, the outflow of the \gls{ST} can be computed as
\begin{equation}
 q_{\textup{st}}(k)=r_{\textup{st}}(k)\textup{min}\bigg[q_{s_{1}}(k-\delta)+\frac{w_{\textup{st}}(k)}{T},q_{\textup{max,st}}(k)\Bigg],
 \label{equation114}
\end{equation}
where $q_{\textup{max,st}}(k)$ indicates the capacity of the \gls{ST}, which is directly influenced by the density of the next link, i.e., $\rho_{s_{2}}(k)$, as follows
\begin{equation}
   q_{\textup{max,st}}(k) =
   \begin{cases}
  Q_{\textup{st}} & \text{if $\rho_{s_{2}}(k)<\rho_{\textup{cr},s_{2}}$} \\
  Q_{\textup{st}}P(k) & \text{otherwise}.
  \label{equation20}
\end{cases}
\end{equation}
Now, it is necessary to prevent the formation of a potential overloaded queue of vehicles at the entrance of the \gls{ST} during each time interval $k$. Inspired by stationary queuing model of traffic flow networks \cite{van2007modeling}, we apply a manipulation of the flow entering the \gls{ST}, i.e., $q_{s_{1}}(k)$, based on the dynamics of the \gls{ST}. Then, let $\ell_{\textup{max,st}}$ denote the maximum number of vehicles that can stop at the \gls{ST}, which is computed according to the length of the \gls{ST} and the average length of a vehicle. Then, the flow entering the \gls{ST}, can be constrained by the space available at the \gls{ST}, i.e.,
\begin{equation}
 q_{s_{1}}(k)=\textup{min}\bigg[\rho_{s_{1}}(k)v_{s_{1}}(k),\frac{\ell_{\textup{max,st}}-\ell_{\textup{st}}(k)}{T}\bigg].
 \label{equation22}
 \end{equation}
 \begin{figure*}
  \includegraphics[width=\textwidth]{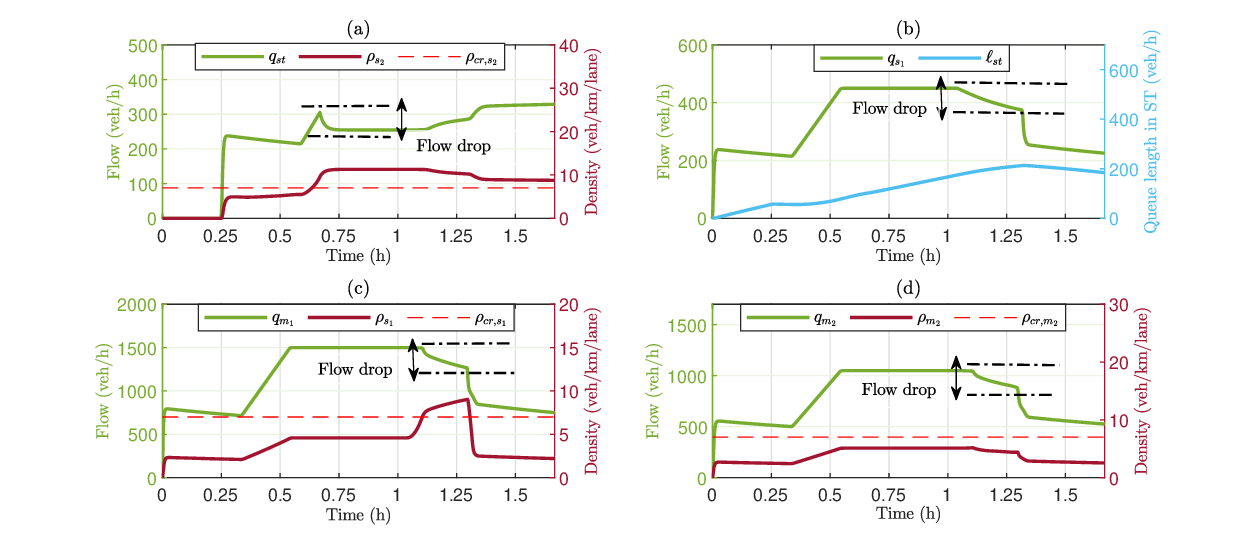}
  \caption{Back propagation phenomenon resulting from congestion, hence $\rho_{s_{2}}>\rho_{\textup{cr},s_{2}}$ at the off-ramp of the \gls{ST} $s_{2}$. (\textup{a}) the flow of \gls{ST} $q_{\textup{st}}(k)$ and  density of $s_{2}$ $\rho_{s_{2}}(k)$, (\textup{b}) flow of on-ramp $s_{1}$ $q_{s_{1}}(k)$, queue length of vehicles in \gls{ST} $\ell_{\textup{st}}(k)$, (\textup{c}) flow of $m_{1}$ $q_{m_{1}}(k)$, density of link $s_{1}$ $\rho_{s_{1}}(k)$, (\textup{d}) the flow and density in $m_{2}$, i.e., $q_{m_{2}}(k)$, $\rho_{m_{2}}(k)$.}  
  \label{backpropagation}
\end{figure*}

 Note that the proposed \gls{METANET-s} has been designed to be easily integrated with various traditional and advanced control strategies aimed at preventing or reducing traffic congestion.
 For example, a ramp-metering mechanism \cite{knoop2018ramp} based on model predictive control approaches can be designed to regulate the flow of vehicles merging back into the mainstream, i.e., $q_{\mathrm{st}}(k)$. This control policy is grounded in the idea of reducing the overall traffic congestion by reducing the number of  vehicles entering the main stream during peak hours. Another possibility is to design incentive-based traffic control policies to incentivize drivers to exit the main stream and stop at the \gls{ST} to take advantage of services at discounted prices, such as charging electric vehicles at a cheaper price~\cite{cenedese2021highway}. In such a framework, game-theoretical approaches can be used to design control policies for $\gamma_{m_1,s_1}(k)$ and $\delta$.

\section{Simulation results}

We consider a freeway stretch consisting of $N=11$ links organized as in Figure~\ref{networksample}, hence $\bs N\coloneqq\{o,\tup{st},s_1,s_2,m_1,\dots,m_7\}$. The origin link $o\in\N$ is entry point for the incoming demand $d_{o}(k)$, while \{$m_{1},\dots,m_{7}$\} describe the mainstream. The traffic condition after the destination link $m_{7}$ is assumed to be uncongested. Furthermore, the length of the \gls{ST} is $1$\,\textup{km}, while the length of the links $\{o,s_1,s_2,m_1,\dots,m_7\}$ are equal to $0.3$\,\textup{km} with $\lambda=3$ lanes. 
The links $\{s_1,\tup{st},s_2\}$ describe the \gls{ST}, where $s_1$ and $s_2$ are the off and on-ramps respectively, while $\tup{st}$ is the \gls{saf} link modeling the time spent at the \gls{ST}. 
The time interval in the simulations is $T=0.01$\,\textup{s} and we simulate a total of $1.6$\,\textup{h}, hence $k\in[0,6\cdot10^4]$.  
To mimic a classic peak hour demand,
$d_{o}(k)$ is assumed to be the following piece-wise linear demand function
\begin{equation} \label{equation26}
  d_{o}(k)=\max\left(500,-\tfrac{1}{9}|k-30000|+2500 \right).    
\end{equation}
Of the total demand, $30\%$ of it stops at the \gls{ST}, and the remainder $70\%$ remains on the mainstream. Table \ref{tab2} contains the static traffic parameters used for the dynamics.

\begin{table}
\caption{The traffic parameters used in the simulations of the \gls{METANET-s} model}
\label{tab2}
\begin{center}
 \begin{tabular}{|c|c|c|c|c|c|c|}
 \hline
 $\tau$   & $\bar v_{m}$ & $\rho_{\textup{max}}$ & $K$ & $V$ & $\rho_{\textup{cr}}$  &  $a_{m}$ \\
 \hline
 \hline
  $0.005$ &   $102$ & $30$ & $40$ & $60$&  $20$ &  $2.34$\\
  \hline
 \end{tabular}   
\end{center}  
\end{table}

\subsection{Traffic congestion back propagation in \gls{METANET-s}}
Here, we show that \gls{METANET-s} can describe the back propagation of a traffic congestion. 
As shown in Figure~\ref{backpropagation}.a, at $t=0.7$\,\textup{h} a congestion occurs in the link $s_{2}$, hence $\rho_{s_{2}}(k)>\rho_{\textup{cr},s_{2}}$. This congestion leads to a decrease in the available capacity of the \gls{ST} $q_{\textup{max,st}}(k)$ as per \eqref{equation20}. As a result, the number of vehicles at the \gls{ST} $\ell_{\textup{st}}(k)$ increases, see Figure~\ref{backpropagation}.b. 
Then the saturation in \eqref{equation22} activates to limit $q_{s_1}(k)$. Due to the reduced capacity of the \gls{ST} $q_{\textup{max,st}}(k)$, the total number of vehicles in the \gls{ST} $\ell_{\textup{st}}(k)$ %\CC{\tt I thought this was the total number of vehicles not the queuing ones.} 
peaks at $1.3$ \textup{h}, showing a significant backlog of vehicles in $s_{1}$ waiting to enter to the \gls{ST}, see $\rho_{s_1}(k)$ in Figure~\ref{backpropagation}.c. 

\begin{figure*}
  \includegraphics[width=\textwidth]{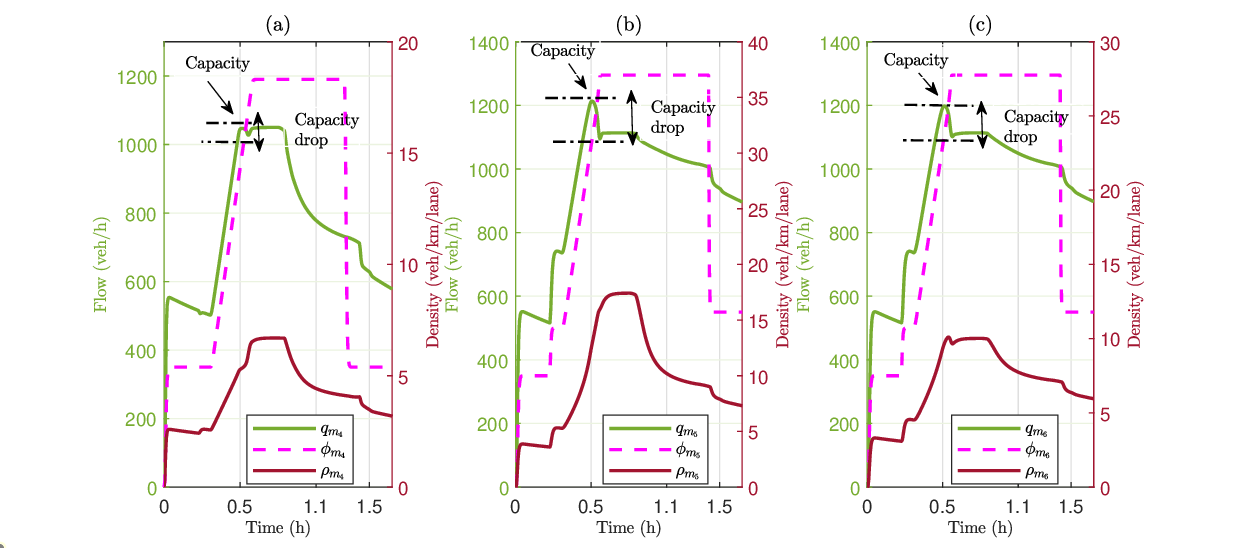}
  \caption{Capacity drop captured in the \gls{METANET-s} in comparison with the flow variations captured by the model of \gls{CTM-s} at On-ramp merging location: flow and density trajectories for links $m_{4}$,$m_{5}$, and $m_{6}$ in \gls{METANET-s} and flow captured by \gls{CTM-s} ($\phi(k)$) of cells $m_{4}$,$m_{5}$, and $m_{6}$ respectively in (\textup{a}), (\textup{b}), and (\textup{c}).}
  %\CC{Update the caption.}}
  \label{Capacity drop}
\end{figure*}

\subsection{Capacity drop in \gls{METANET-s} comparison with \gls{CTM-s}}
 
We compare now the ability of \gls{METANET-s} to describe capacity drops compared to the \gls{CTM-s}.
In Figure~\ref{networksample}, the flow and density of links $m_{4}$, $m_{5}$, and $m_{6}$ are represented.
Initially, $q_{m_5}(k)$ peaks around 1213\,\textup{veh/h lane} at approximately $0.5$ \textup{h}. Then, there is a significant capacity drop, making it fall at around 1100 \textup{veh/h lane}, hence a 9.5\% reduction, see Figure~\ref{Capacity drop}.b. As shown in Figure~\ref{Capacity drop}.c the evolution of $q_{m_6}(k)$ is similar to $q_{m_5}(k)$.

In Figure \ref{Capacity drop}.a, we plot $q_{m_4}(k)$. Due to the propagation of a shock wave, there is a drop in traffic flow in link $m_4$. This drop is accompanied by a rapid increase in density, indicating that congestion was forming as vehicles approached the bottleneck.

Noticeably, simulation results in Figures~\ref{Capacity drop}.a-c show that \gls{CTM-s}, implemented as in \cite{cenedese2022novel}, is not able to capture the presence of the capacity drop and highly overestimates the flow among links. This behavior is consistent with our other numerical simulation allowing us to conclude that \gls{METANET-s} has better-predicting capabilities than \gls{CTM-s} in case of congestion.

%%%%%%%%%%%%%%%%%%%%%%%%%%%%%%%%%%%%%%%%%%%%%%%%%%%%%%%%%%%%%%%%%%%%%%%%%%%%%%%%
\section{Conclusion}
Inspired by the classical METANET model, in this paper we have developed a novel traffic model: the \gls{METANET-s}. It is a second-order macroscopic traffic model incorporating service stations dynamics. The capability of the \gls{METANET-s} to model speed dynamics excels in capturing intricate traffic phenomena, such as capacity drops, thereby addressing the limitations of the recently proposed \gls{CTM-s}. Including a constraint on the flow entering the \gls{ST} allows us to take into account the \gls{ST} capacity. Furthermore, the \gls{METANET-s} has been developed to aid in the design of innovative control strategies aimed to incentivize drivers to stop at  \glspl{ST} to prevent and/or reduce traffic congestion.  
An interesting direction is to use tools from feedback optimization to control the on and off-ramps of the \gls{ST} this can allow us to take advantage of the more precise dynamics proposed while overcoming the challenges due to the highly nonlinear dynamics.

\balance
\bibliographystyle{IEEEtran}
\bibliography{ref}

% \printglossaries
\end{document}